\documentstyle[twocolumn,seceq,epsf]{jpsj}
\def\runtitle{Two-Particle Self-Consistent Theory for the Single-Impurity Anderson Model}
\def\runauthor{Tetsuro {\sc Saso}}

\def\D{\Delta}
\def\e{\epsilon}
\def\G{\Gamma}
\def\w{\omega}
\def\r{\rho}

\def\s{\sigma}
\def\d{\mbox{d}}
\def\i{\mbox{i}}

\def\i{{\mbox{\scriptsize i}}}

\def\dd{{\mbox{\scriptsize d}}}

\def\bra{\langle}
\def\ket{\rangle}

\def\lsim{$\raisebox{-0.5ex}{$\stackrel{<}{\sim}$}$}

\title
{
Investigation of the Two-Particle-Self-Consistent Theory for the Single-Impurity Anderson Model and an Extension to the Case of Strong Correlation}
\author
{
Tetsuro {\sc Saso}\footnote{E-mail: saso@phy.saitama-u.ac.jp}
}
\inst
{
Department of Physics, Faculty of Sciences, Saitama University, Urawa 338-8570
}
\recdate
{
\hspace{4cm}
}

\abst
{
The two-particle-self-consistent theory is applied to the single-impurity Anderson model.  It is found that it cannot reproduce the small energy scale in the strong correlation limit.  A modified scheme to overcome this difficulty is proposed by introducing an appropriate vertex correction explicitly.  Using the same vertex correction, the self-energy is investigated, and it is found that under certain assumptions it reproduces the result of the modified perturbation theory which interpolates the weak and the strong correlation limits.
}
\kword
{Impurity Anderson model, two-particle self-consistent theory, dynamical susceptibility, self-energy
}

\begin{document}
\sloppy
\maketitle

\section{Introduction}
The single impurity Anderson model (SIAM)\cite{Anderson61} is not only a model for the magnetic impurity in metals, chemisorption\cite{Newns69} and a quantum dot in mesoscopic systems,\cite{Goldhaber-Gordon98} but also the most important model for the study of the strongly correlated electron systems.
It was proved that the lattice models with strong correlation is reduced to a generalized impurity Anderson model with energy-dependent hybridization in the limit of infinite spatial dimensions.\cite{Georges96}
In that theory, a knowledge on the excitation spectrum of the impurity is indispensable.

It is, however, not possible to obtain an exact solution to the dynamical properties of SIAM.  The best way known to date is to resort to the numerical renormalization group method,\cite{Yoshida90} but
it needs heavy numerical tasks, especially when both finite temperature, finite magnetic field and finite excitation energy must be taken into account.  Therefore, it is desirable to find an approximate but a simpler scheme which is applicable to the problems in realistic materials with strong correlation.

Various approximate schemes proposed to date have several deficiencies despite their advantages.  The non-crossing approximation (NCA)\cite{Kuramoto83,Grewe83,Coleman84} cannot reproduce the Fermi liquid properties at low frequencies and low temperatures, and the slave boson mean field or $1/N$ expansion theory\cite{Read83} cannot avoid redundant phase transitions.
The second-order perturbation theory exhibits Fermi liquid properties and happens to reproduce the correct atomic limit in the symmetric case\cite{Yamada75}, but it is not the case when there is an electron-hole asymmetry.
This point is recently overcome by introducing an interpolative form for the self-energy phenomenologically.\cite{Martin-Rodero82}

Therefore, it is of much importance to establish a method to calculate the static and dynamic properties of SIAM in the wide range of parameters efficiently.
Such a method must be free from the Hartree-Fock instability and have proper energy scales.
In the present paper, we first apply the two-particle self-consistent theory (TPSC)\cite{Vilk94} to SIAM in \S 2.  TPSC has proved to be successful in the study of the Hubbard model in the weak and the intermediate correlations, and is considered to be superior to the fluctuation-exchange approximation (FLEX)\cite{Bickers91} in that only TPSC can reproduce the side peaks in the single-particle spectrum.
Application of FLEX to SIAM is already reported\cite{White92}, which has clarified that FLEX is poor on magnetic susceptibility than the simple perturbation theory,\cite{Horvatic87} in addition to the above-mentioned deficiency on the spectrum.
Despite the superiority of TPSC, we will find below that it cannot reproduce a small energy scale in SIAM in the strong correlation limit.

TPSC can be also regarded as an extension of the self-consistently renormalized spin-fluctuation theory (SCR).\cite{Moriya85}  The author previously proposed an extension of SCR theory\cite{Moriya95} to the case of the strong correlation on a microscopic basis.\cite{Saso99}  In \S 3 we apply a similar scheme to SIAM in combination with TPSC instead of SCR, and establish a scheme which can be applied to the case of strong correlation.  We also investigate the self-energy using the same vertex corrections and find that under certain assumptions our theory reproduces the modified perturbation theory (MPT),\cite{Martin-Rodero82,Levy-Yeyati93,Takagi99} which interpolates the weak and strong correlation limits.  Thus the phenomenological character of MPT is partially resolved.

\section{The Two-Particle-Self-Consistent Theory for the Single-Impurity Anderson Model}
The Hamiltonian for SIAM is written as
\begin{eqnarray}
  {\cal H} &=& \sum_{k\s} \e_kc^+_{k\s}c_{k\s} + E_{\dd} \sum_{\s} n_{d\s} \nonumber \\
   & & + V\sum_{k\s} (d^+_{\s}c_{k\s} + c^+_{k\s}d_{\s}) + U n_{d\uparrow} n_{d\downarrow}
\end{eqnarray}
in the ordinary notation, where $\e_k$ and $E_{\dd}$ denote the energy of the conduction and d electrons, respectively, $V$  the hybridization, and $U$ the Coulomb repulsion between d-electrons.  In the following, the subscript d in $n_{d\s}$ will be dropped.

In TPSC, the charge and the spin susceptibilities are expressed in the form similar to those of RPA, but with the renormalized interaction parameter $U_c$ and $U_s$ for charge and spin channels, respectively.
These parameters are determined by the sum rules,
\begin{eqnarray}
  T\sum_{\w} \frac{2\Pi_0(\i\w)}{1+U_c\Pi_0(\i\w)} &=& n + 2\langle n_{\uparrow}n_{\downarrow}\rangle -n^2, \label{eq:sumrule1}\\
  T\sum_{\w} \frac{2\Pi_0(\i\w)}{1-U_s\Pi_0(\i\w)} &=& n - 2\langle n_{\uparrow}n_{\downarrow}\rangle. \label{eq:sumrule2}
\end{eqnarray}
Here $\w$ denotes the Matsubara frequency, $n=\langle n_{\uparrow}\rangle+\langle n_{\downarrow}\rangle$ and $U_s$ are related to $\langle n_{\uparrow}n_{\downarrow}\rangle$ by $U_s=U\langle n_{\uparrow}n_{\downarrow}\rangle /\langle n_{\uparrow}\rangle\langle n_{\downarrow}\rangle$.
Thus, $U_s$ can be determined from the second equation, while $U_c$ can be obtained by solving the first equation once $U_s$ is solved.
We assume the symmetric case $E_d=-U/2$ and the paramagnetic state $\langle n_\s\rangle=n/2$ throughout the present paper.  An extension to the asymmetric case is straightforward.
$\Pi_0(\w)$ denotes the polarization function for $U=0$, which is calculated at $T=0$ as\cite{Kuramoto90}
\begin{equation}
  \Pi_0(\w)=\frac{2\D}{\pi\w(\w+2\i\D)}\log\left(1-\frac{\i\w}{\D}\right),
\end{equation}
where $\D$ denotes the resonance width of the impurity, $\D=\pi \r_c V^2$ ($\r_c$ is the density of states of conduction electrons at the Fermi energy).
At low frequencies, it behaves as
\begin{equation}
  \Pi_0(\w) \simeq \frac{\Pi_0}{1-\i\w/\D} \qquad (\w \ll \D), \label{eq:Pi0}
\end{equation}
with $\Pi_0=1/\pi\D$.  Using this form, the sum rule eq.(\ref{eq:sumrule2}) reads
\begin{equation}
  1-\frac{U_s}{2U} \simeq \frac{2}{\pi^2} \log \frac{\w_c}{\G},
\end{equation}
where $\w_c$ denotes a cutoff frequency of the order $O(\D)$ and $\G\equiv\D-U_s/\pi$.

For $U\rightarrow \infty$, we have $\langle n_{\uparrow}n_{\downarrow}\rangle \rightarrow 0$ and hence $U_s \rightarrow 0$.
Then the above equation is solved as $\G=\w_c \exp(-\pi^2/2) \approx \w_c/139$.
Namely, we obtain
\begin{equation}
U_s=\pi[\D-\w_c e^{-\pi^2/2}].
\end{equation}
If we set $g\mu_B=1$ ($g=2$ is the $g$-factor), the static susceptibility $\chi\equiv\chi^{zz}$ is given by
\begin{equation}
\chi(\w=0)=\frac{1}{2}\frac{\Pi_0}{1-U_s\Pi_0}=\frac{1}{2\pi\G}.
\end{equation}
Therefore, if we define the Kondo temperature $T_{\mbox{\small K}}$ by $\chi(\w=0)=1/4T_{\mbox{\small K}}$,
we obtain $T_{\mbox{\small K}}=(\pi/2)\G=(\pi/2)\w_c \exp(-\pi^2/2)$, which is finite even for $U\rightarrow \infty$.  Of course, the correct formula is $T_{\mbox{\small K}}=\sqrt{U\D/2}\exp(-\pi U/8\D)$,\cite{Okiji83} which vanishes for $U \rightarrow \infty$.
For $U < \infty$, a similar analysis yields $T_{\mbox{\small K}}(U)=(\pi/2)\w_c \exp[-(\pi^2/2)(1-U_s/2U)]$, which is larger by $\exp[(\pi^2/2)(U_s/2U)]$ than $T_{\mbox{\small K}}(U=\infty)$.

Thus, in TPSC, the renormalized interaction $U_s$ is reduced at $U\gg \D$ so as to avoid the magnetic instability, but does not reach $\pi\D$, namely, $U_s \rightarrow \pi[\D - \w_c\exp(-\pi^2/2)]$ and hence the Kondo temperature remains finite even at $U=\infty$.
Therefore, TPSC cannot describe the case of strong correlation properly.
We will seek for an improvement of TPSC in the next section.  Before it, however, we briefly discuss on the self-energy in TPSC.

The self-energy for the d-electrons is calculated as\cite{Vilk94}
\begin{eqnarray}
  \Sigma_{\s}(\i\e) &=& Un_{-\s} + \frac{U}{4}T\sum_\w \left[ \frac{2U_s\Pi_0(\i\w)}{1-U_s\Pi_0(\i\w)}\right. \nonumber \\
    & & + \left.\frac{2U_c\Pi_0(\i\w)}{1+U_c\Pi_0(\i\w)} \right]G_0(\i\e+\i\w).
\end{eqnarray}
In the weak coupling limit $U\rightarrow 0$, this is reduced to the second-order self-energy
\begin{equation}
  \Sigma_{\s}(\i\e) = Un_{-\s} + U^2T\sum_\w \Pi_0(\i\w)G_0(\i\e+\i\w).
\end{equation}
It is well known, however, that the second-order self-energy does not have an expected spectral properties in the electron-hole asymmetric cases.  For example, the positions of the $E_d$ and $E_d+U$ peaks are not correct, and the Friedel sum rule is not satisfied.
These difficulties can be most easily overcome by the use of the interpolative form of the self-energy\cite{Martin-Rodero82} and by subtracting $\Sigma(0)$ from $\Sigma(\e)$ as pointed out by Yamada.\cite{Yamada79}  More improved treatment is possible by introducing the effective d-level position.\cite{Levy-Yeyati93, Takagi99}

\section{Improvement of Two-Particle-Self-Consistent Theory}
\begin{figure}
\vspace{1cm}
\epsfxsize=7cm
\centerline{\epsfbox{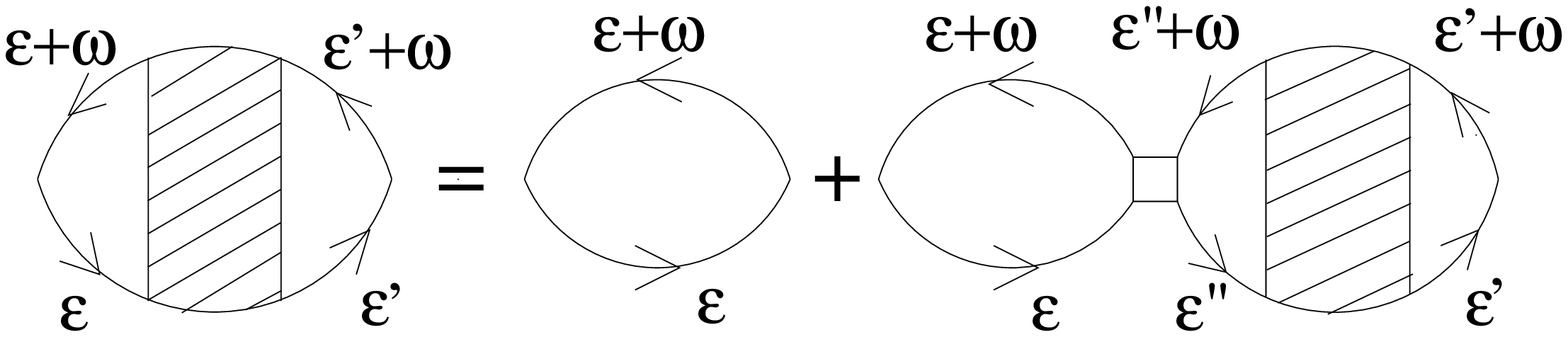}}
\caption{The Feynman diagram for the susceptibility $\chi^{+-}(\i\e,\i\e',\i\w)$. The square denotes the vertex function $\G(\i\e,\i\e',\i\w)$.}
\label{fig:chi}
\end{figure}

The dynamical susceptibility of the Anderson impurity can be generally expressed by the diagram shown in Fig.\ref{fig:chi} and the equation,
\begin{equation}
  \chi^{+-}(\i\w)\equiv T \sum_{\e\e'}\chi^{+-}(\i\e,\i\e',\i\w),
\end{equation}
\begin{eqnarray}
  & & \chi^{+-}(\i\e,\i\e',\i\w)=-G_{\uparrow}(\i\e)G_{\downarrow}(\i\e'+\i\w) \nonumber \\
  & & \times \left[ \delta_{\e,\e'}+T\sum_{\e''}\Gamma(\i\e'',\i\e',\i\w)\chi^{+-}(\i\e'',\i\e',\i\w) \right].
\end{eqnarray}
\begin{figure}
\vspace{0.5cm}
\epsfxsize=7cm
\centerline{\epsfbox{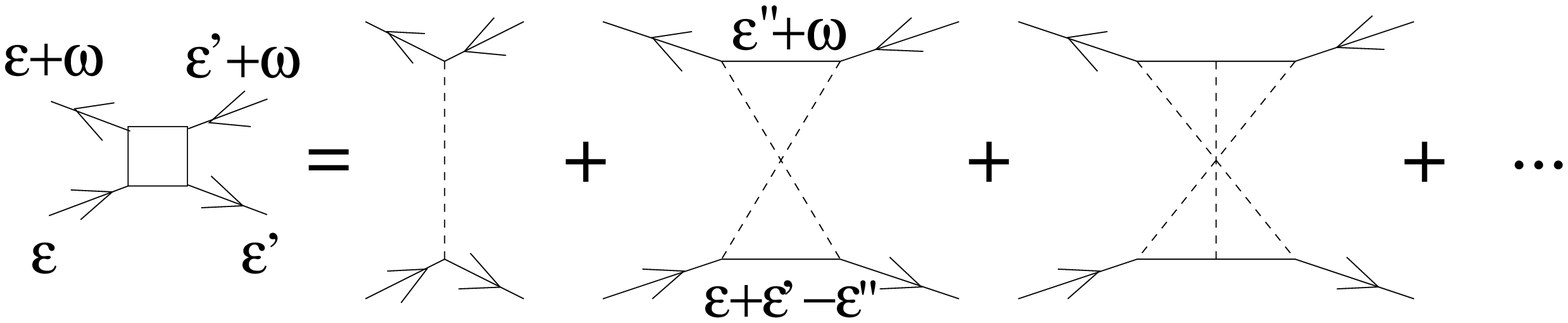}}
\caption{The diagram for the vertex function $\G(\i\e,\i\e',\i\w)$.}
\label{fig:vertex}
\end{figure}
We calculate the irreducible vertex function $\Gamma(\i\e,\i\e',\i\w)$ approximately by taking account of the maximally crossed diagrams (Fig.\ref{fig:vertex}) as
\begin{equation}
  \G(\i\e,\i\e',\i\w)=\frac{U}{1+UK(\i\e+\i\e',\i\w)},
\end{equation}
\begin{equation}
  K(\i\e+\i\e',\i\w)=T\sum_{\e''} G_{\downarrow}(\i\e+\i\e'-\i\e'')G_{\uparrow}(\i\e''+\i\w).
\end{equation}
We use the unperturbed Green's function $G_{0\s}$ for $G_{\s}$ and denote $K$ by $K_0$.  Then, we find that $K_0(0,\w)=\Pi_0(\w)$, $K_0(\w,0)=\Pi_0(\w)$ (note that $\e+\e'$ is bosonic) and thus at low frequencies
\begin{equation}
  K_0(0,\w) = K_0(\w,0) \simeq \frac{1}{\pi\D}\frac{1}{1-\i\w/\D}.
\end{equation}
If we approximate $\G(\i\e,\i\e',\i\w)$ by $\G(\i\w) \equiv \G(0,0,\i\w)$, then we obtain
\begin{equation}
  \chi^{+-}(\i\w)=\frac{\Pi_0(\i\w)}{1-\G(\i\w)\Pi_0(\i\w)}, \label{eq:chi}
\end{equation}
\begin{equation}
\G(\i\w)=\frac{U}{1+U\Pi_0(\i\w)}.
\end{equation}
Note that at $T=0$ and the low frequency limit, the effective interaction becomes
\begin{equation}
\G(0)=\frac{U}{1+U/\pi\D},
\end{equation}
which smoothly interpolates between $\G(0)=U$ for small $U$ and $\G(0) \rightarrow \pi\D$ for $U \rightarrow \infty$.  Therefore, the magnetic susceptibility $\chi^{zz}(0)=(1/2)\chi^{+-}(0)$ diverges when and only when $U \rightarrow \infty$.  This is a desired property.  Within the present approximation, eq.(\ref{eq:chi})  becomes
\begin{equation}
  \chi^{+-}(\i\w)=\Pi_0(\i\w)[1+U\Pi_0(\i\w)]. \label{eq:chi2}
\end{equation}
Namely, the terms higher than the second order with respect to $U$ in the perturbational expansion of $\chi(\w)$ are completely canceled out with the terms from the vertex correction, which is of course an artifact of the approximation.  Note also that eq.(\ref{eq:chi2}) is correct up to $O(U)$.\cite{Yamada79,Saso99}

\begin{figure}
\vspace{1cm}
\epsfxsize=6cm
\centerline{\epsfbox{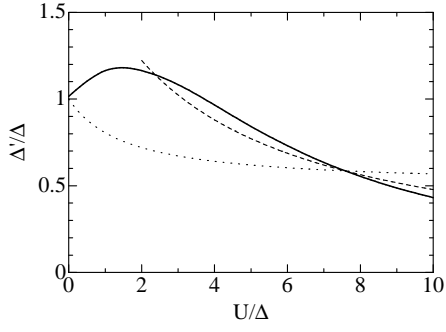}}
\caption{$\D'/\D$ calculated by eq.(\ref{eq:Delta}) is shown as a function of $U$.  Also shown are $\D'/\D = 2\pi\D/(\pi\D+U)$ (dashed line) and $\tilde{\D}/\D = 1/[1+U/(\pi\D+U)]$ (dotted line).}
\label{fig:Delta-U}
\end{figure}

\begin{figure}
\vspace{1cm}
\epsfxsize=6cm
\centerline{\epsfbox{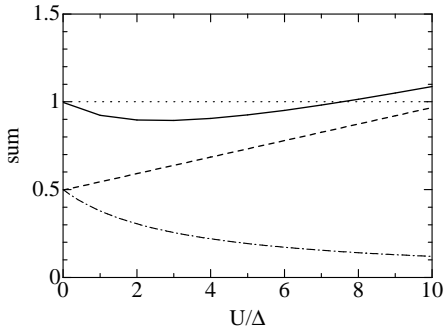}}
\caption{The left-hand side of eq.(\ref{eq:sum}) is plotted by the full line for the case without the $C$ term correction.  The first term in l.h.s. is plotted by the dashed line, whereas the second term by the dash-dotted line.}
\label{fig:sum}
\end{figure}

\begin{figure}
\vspace{1cm}
\epsfxsize=7cm
\centerline{\epsfbox{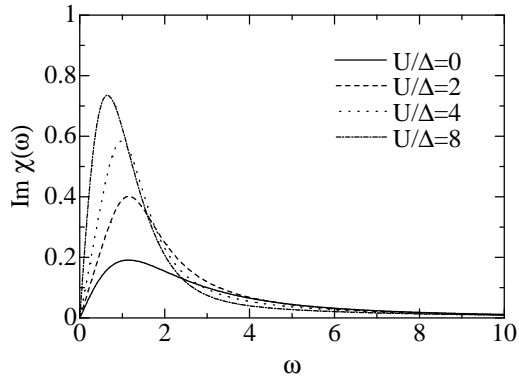}}
\caption{The imaginary parts of the dynamical susceptibility $\chi^{+-}(\w)'$ are shown for $U/\D=$0, 2, 4 and 8.}
\label{fig:chi(w)}
\end{figure}

Using the low frequency form for $\Pi_0(\w)$, we obtain
\begin{equation}
  \chi^{+-}(\w) \simeq \frac{2\tilde{\chi}_0}{1-\i\w/\tilde{\D}},
\end{equation}
where $\tilde{\chi}_0=(1/2\pi\D)(\pi\D+U)/\pi\D$ and $\tilde{\D}=\D/[1+U/(\pi\D+U)]$.
On increasing $U$ from 0 to $\infty$, $\tilde{\chi}_0$ varies from $1/(2\pi\D)$ to $U/2(\pi\D)^2 \rightarrow \infty$.
Thus the Kondo temperature $T_{\mbox{\small K}}=1/4\tilde{\chi}_0$ vanishes for $U\rightarrow \infty$ as we desired, but more slowly as $T_{\mbox{\small K}} \rightarrow (\pi\D)^2/2U$ than the correct behavior.  This must be improved in the future by taking account of better vertex corrections.
On the contrary, $\tilde{\D}$ is reduced from $\D$ only to $\D/2$ for $U=0 \rightarrow \infty$, and remains finite at $U = \infty$.
Therefore, we replace $\tilde{\D}$ with an effective parameter $\D'$ and determine it so as to satisfy the sum rule eq.(\ref{eq:sumrule2}) with $U_s=\G(0)$.  Namely, in the symmetric case and at $T=0$,
\begin{eqnarray}
  1-\frac{\G(0)}{2U} &=& \frac{1}{\pi} \int_0^{\w_c} \d\w\ \mbox{Im} \frac{4\tilde{\chi}_0}{1-\i\w/\D'} \nonumber \\
  &=& \frac{4\tilde{\chi}_0\D'}{\pi}\log\frac{\w_c}{\D'}.\label{eq:311}
\end{eqnarray}
For $U=0$, we have $\G(0)=0$, and the cutoff $\w_c$ should be chosen as $\w_c=\D\exp(\pi^2/4)$ since $\tilde{\chi}_0=1/(2\pi\D)$ and $\D'=\D$.  For $U>0$, the cutoff may also be chosen as a similar form $\w_c=\D'\exp(\pi^2/4)$. Then we obtain $\D' = \D\times 2\pi\D/(\pi\D+U)$, which vanishes as $U$ increases.  If we use eq.(\ref{eq:Pi0}) for $\Pi_0(\w)$, the integration converges without a cutoff.
In that case, we previously proposed\cite{Saso99} to modify $\chi^{+-}(\w)$ into
\begin{equation}
  \chi^{+-}(\w)'=\frac{1}{\chi^{+-}(\w)^{-1}-\i C\w},
\end{equation}
and determine the parameter $C$ by the sum rule.  This is essentially corresponding to modifying $\tilde{\D}$ as mentioned above.  At low frequencies, $\chi^{+-}(\w)'$ behaves as
\begin{equation}
  \chi^{+-}(\w)' \simeq \frac{2\tilde{\chi}_0}{1-\i\w/\D'},
\end{equation}
where
\begin{equation}
  \D'=(\tilde{\D}^{-1}+2C\tilde{\chi}_0)^{-1} \label{eq:Delta}
\end{equation}
which tends to $\D (\pi/C)\pi\D/(\pi\D+U) \rightarrow 0$ for $U \rightarrow \infty$.  In Fig.\ref{fig:Delta-U}, we show this $\D'$ and compare it with $\D' = \D\times 2\pi\D/(\pi\D+U)$ in the above analysis using eq.(\ref{eq:311}).  Both are in good agreement for $U \gg \D$ and become smaller than $\tilde{\D}$ for $U/\D > 7$ although the reduction of $\D'$ is not sufficient quantitatively.
The first rise of $\D'$ at $U \lsim \D$ is to compensate the decrease of the left-hand side of the sum rule
\begin{equation}
\frac{1}{\pi} \int_0^\infty \! \dd\w\ \mbox{Im}\frac{2\Pi_0(\w)}{1-\G(\w)\Pi_0(\w)}+\frac{\G(0)}{2U} = 1 \label{eq:sum}
\end{equation}
from 1 to 0.9 for $U/\D=0$ to 2 when the $C$ term is not included (see Fig.\ref{fig:sum}). This decrease is traced back to the decrease of $\G(0)/2U=(1/2)/(1+U/\pi\D)$.  For $U/\D>2$, the increase of the first term of the l.h.s. of the above equation dominates and l.h.s. increases again.
The imaginary parts of the dynamical susceptibility calculated with this $C$ term correction are shown in Fig.\ref{fig:chi(w)}.  We set $\D=1$ here and henceforth.

\begin{figure}
\vspace{1cm}
\epsfxsize=7cm
\centerline{\epsfbox{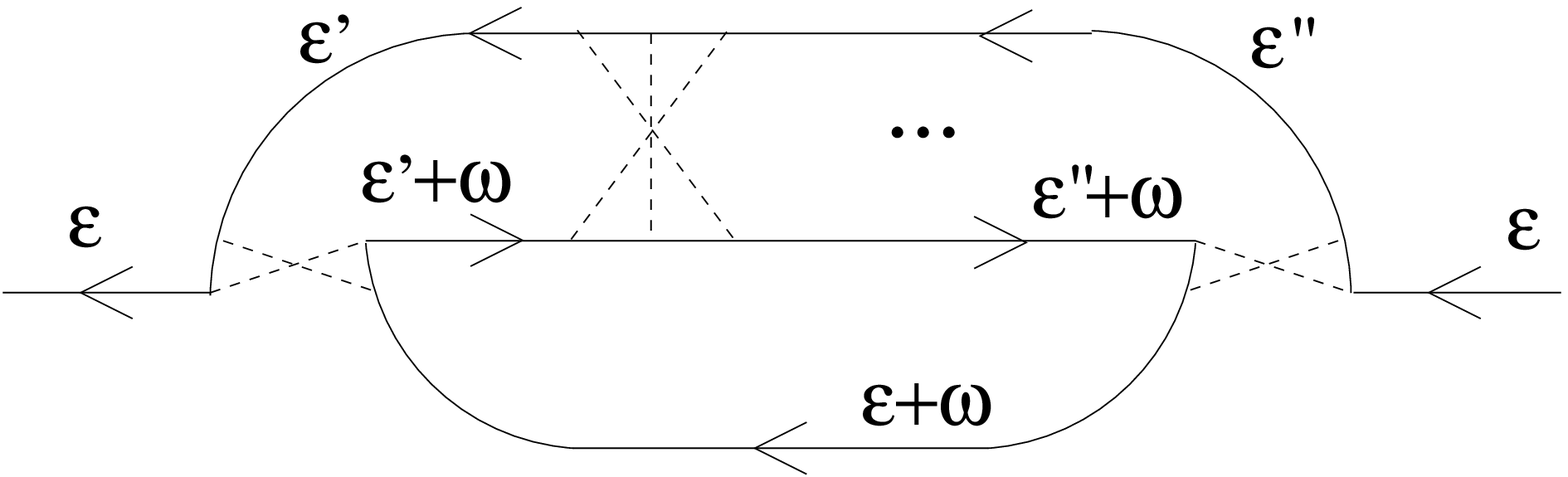}}
\vspace{1cm}
\epsfxsize=7cm
\centerline{\epsfbox{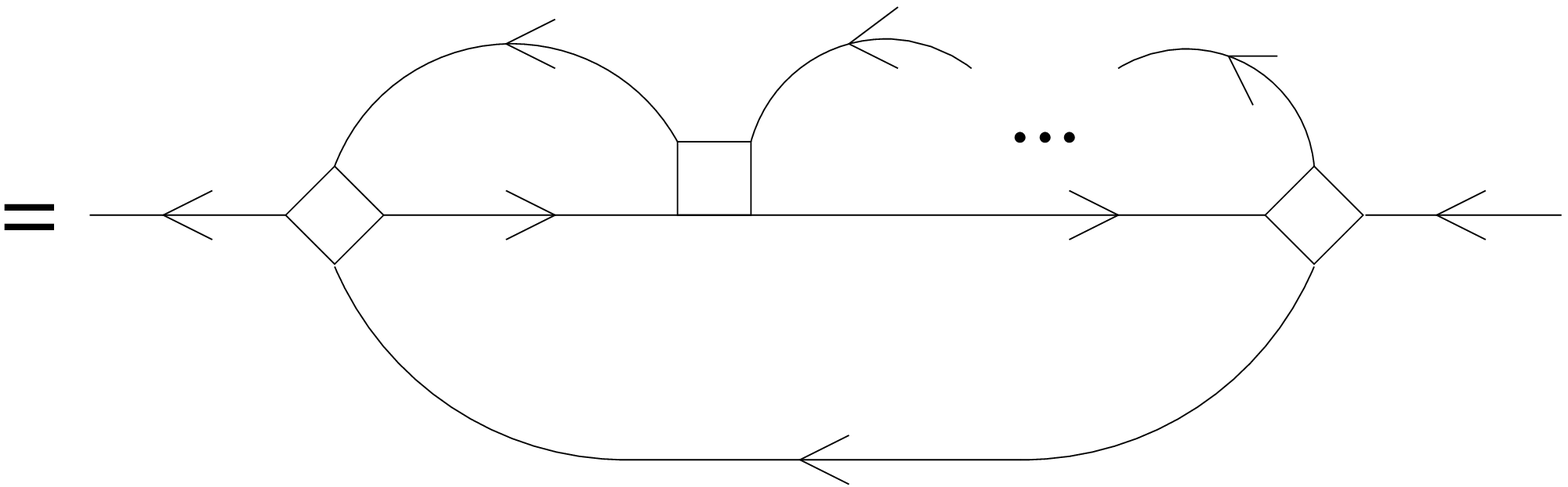}}
\caption{A typical Feynman diagram for the self-energy.}
\label{fig:self}
\end{figure}

\begin{figure}
\vspace{1cm}
\epsfxsize=6cm
\centerline{\epsfbox{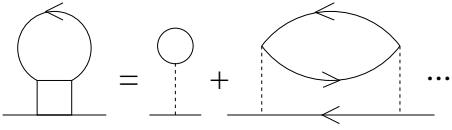}}
\caption{The left- and right-hand sides denote eqs.(\ref{eq:1st2nd-a}) and (\ref{eq:1st2nd-b}), respectively.}
\label{fig:1st2nd}
\end{figure}

Next, we investigate the self-energy. The Feynman diagram is shown in Fig.\ref{fig:self}, which may be expressed as
\begin{eqnarray}
  \Sigma_{\uparrow}(\i\e) &=& Un_{\downarrow} + \cdots \nonumber \\
  &+& T^3\sum_{\e',\e'',\w,\cdots}
   \Gamma(\i\e,\i\e',\i\w) G_{\uparrow}(\i\e')G_{\downarrow}(\i\e'+\i\w) \nonumber \\
   &\times& \Gamma(\i\e',\i\e'',\i\w) \cdots G_{\uparrow}(\i\e'')G_{\downarrow}(\i\e''+\i\w) \nonumber \\
   & \times &  \Gamma(\i\e'',\i\e,\i\w)
   \times G_{\downarrow}(\i\e+\i\w) + \cdots.
\end{eqnarray}
Keeping the behaviors of $K_0(0,\w)$ and $K_0(\e,0)$ in mind, we approximate $\Gamma(\i\e,\i\e',\i\w)$ at the left and the right ends by the separable form $[\bar{\Gamma}'(\i\e)\Gamma(\i\w)]^{1/2}$,where $\bar{\G}'(\i\e)$ denotes some average of $\Gamma(\i\e,\i\e',\i\w)$ on $\e'$ and $\w$.  To fix $\bar{\G}'(\i\e)$, we consider the first order self-energy due to $\Gamma(\i\e,\i\e',\i\w)$ (Fig.\ref{fig:1st2nd}):
\begin{equation}
  \Sigma_{\uparrow}(\i\e) = T\sum_\w \Gamma(\i\e,\i\e,\i\w) G_{\downarrow}(\i\e+\i\w),
\end{equation}
which is expanded for small $U$ as
\begin{eqnarray}
   &=& T\sum_\w \left[ U - U^2 K_0(\i\e+\i\e,\i\w)+ \cdots \right] G_{\downarrow}(\i\e+\i\w). \label{eq:Gamma} \nonumber \\
  &=& Un_{\downarrow} + \Sigma^{(2)}_{\uparrow}(\i\e) + \cdots,\label{eq:1st2nd-b}
\end{eqnarray}
where $\Sigma^{(2)}_{\uparrow}(\i\e)$ denotes the second-order self-energy.
On the other hand, if we replace $\Gamma(\i\e,\i\e,\i\w)$ in $\Sigma_{\uparrow}(\i\e)$ with the average $\bar{\Gamma}(\i\e)\equiv \bra\Gamma(\i\e,\i\e,\i\w)\ket_\w$,
\begin{eqnarray}
  \Sigma_{\uparrow}(\i\e) &=& \bar{\G}(\i\e) T\sum_{\w} G_{0\downarrow}(\i\e+\i\w)  \label{eq:1st2nd-a} \\
  &=& Un_{\downarrow} - U^2 n_{\downarrow} \bra K_0(\i\e+\i\e,\i\w) \ket_\w + \cdots. \nonumber \\
\end{eqnarray}
Comparing these two expressions, we obtain
\begin{equation}
  \bra K_0(\i\e+\i\e,\i\w) \ket_\w = -\Sigma^{(2)}_{\uparrow}(\i\e)/U^2n_{\downarrow}.
\end{equation}
Therefore, we approximate $\bar{\Gamma}(\i\e)$ as
\begin{equation}
  \bar{\Gamma}(\i\e) = \frac{U}{1-B'\Sigma^{(2)}_{\uparrow}(\i\e)/Un_{\downarrow}},
\end{equation}
where $B'$ is an adjustable parameter to take account of the effects of the higher order terms.
Approximating $\bar{\G}'(\i\e)$ by $\bar{\G}(\i\e)$ and replacing the intermediate $\Gamma(\i\e,\i\e',\i\w)$ by the previous $\Gamma(\i\w)$, we obtain
\begin{eqnarray}
  & & \Sigma_{\uparrow}(\i\e) \simeq Un_{\downarrow} + \nonumber \\
  & & \hspace{-6mm} \bar{\G}(\i\e) T\sum_{\w} \G(\i\w)\frac{\Pi_0(\i\w)}{1-\G(\i\w)\Pi_0(\i\w)}G_{0\downarrow}(\i\e+\i\w).
\end{eqnarray}
We further note that
\begin{equation}
  \frac{\G(\i\w)}{1-\G(\i\w)\Pi_0(\i\w)} = U,
\end{equation}
which holds within the present approximation.
Therefore we obtain
\begin{eqnarray}
  \Sigma_{\uparrow}(\i\e) &=& Un_{\downarrow} + \frac{U^2 T\sum_{\w} \Pi_0(\i\w)G_{0\downarrow}(\i\e+\i\w)}{1-B'\Sigma^{(2)}_{\uparrow}(\i\e)/Un_{\downarrow}} \nonumber \\
  &=& Un_{\downarrow} + \frac{\Sigma^{(2)}_{\uparrow}(\i\e)}{1-B\Sigma^{(2)}_{\uparrow}(\i\e)}, \label{eq:MPT}
\end{eqnarray}
where $B=B'/Un_{\downarrow}$.  This is exactly the same form as the interpolative self-energy proposed by Martin-Rodero, et al.\cite{Martin-Rodero82}, which bridges between the second-order and the atomic limit self-energies in both the symmetric and the asymmetric cases.
$B$ was determined so as to reproduce the correct self-energy in the atomic limit when $U \rightarrow \infty$.
Namely, one obtains $B=(1-2n_{-\s})/Un_{-\s}(1-n_{-\s})$.

Since the above formula for the self-energy does not satisfy the Friedel sum rule, one has to introduce an effective d level energy to be adjusted to fulfil the sum rule,\cite{Levy-Yeyati93,Takagi99} or to subtract $\Sigma(0)$ from $\Sigma(\e)$\cite{Yamada75} to approximately satisfy it.
Details of the calculations will be found in ref.\citen{Takagi99}.
Note that $B$ vanishes for the symmetric case.
Namely, the self-energy becomes equal to the second-order one due to the cancelation of the higher order terms in the present approximation.
Recently, it was pointed out that the second-order self-energy becomes exact when the width of the conduction band becomes zero.\cite{Lange98}
The cancelation of the higher order terms in our treatment may be related to this finding.

\section{Conclusions}
In this paper, we have applied the two-particle self-consistent approximation to the single-impurity Anderson model, and found that it is not suitable for the description of the strong correlation limit.  The form of the dynamical susceptibility was assumed to be of RPA type in TPSC, and the effective interaction $U_s$ is determined by the sum rule eq.(\ref{eq:sumrule2}).  In this scheme, a magnetic instability was avoided but $U_s$ remained to be too small in $U \rightarrow \infty$.  Thus, the low energy scale remains finite and does not vanish even at this limit.  This is understandable since TPSC was devised originally for the cases of weak and intermediate correlations.
We needed an improved treatment of the vertex corrections for the dynamical susceptibility.

Then, we have proposed a modification of TPSC applicable to the strong correlation.  We have used the renormalized interaction $\G(\w)=U/[1+U\Pi_0(\w)]$ instead of $U_s$, and assumed the modified form of $\chi(\w)$ as
\begin{equation}
  \chi^{+-}(\w)=\frac{2\Pi_0(\w)}{1-\G(\w)\Pi_0(\w)-2\i C\w\Pi_0(\w)},
\end{equation}
determining the parameter $C$ by the same sum rule.
This scheme has realized the vanishment of the energy scale in the strong correlation limit, although the decrease of the energy scale is slow compared to the correct behavior (Kondo temperature).
Apparently, this is due to our use of the approximate vertex corrections to $\chi(\w)$ which is not exactly consistent with the self-energy diagram.
It is known that the calculation of the static susceptibility by the numerical differentiation of the electron numbers $n_{\uparrow}-n_{\downarrow}$ within MPT leads to a rather correct values if it is combined with the Friedel sum rule.\cite{Takagi99}  But one cannot apply it to the calculation of the dynamics.

We have also investigated the self-energy within the same vertex correction and found that with an appropriate approximation it reproduces the so-called modified perturbation theory which interpolates the perturbative regime and the atomic limit.  Although the present derivation includes a crude approximation, it suggests that an appropriate treatment of the vertex correction may lead to a theory which bridges the weak and strong correlation limits.
Extension of the present scheme to the asymmetric case is straightforward.
In order to improve the present theory further, we need a better vertex function.  It would be of much interest to calculate the set of eqs.(3.1-4) exactly.

We have already applied a theory, which is similar to the present one but is combined with SCR theory, to the description of the quantum critical phenomena.\cite{Saso99}
It will be improved by the application of the present formulation to lattice problems.  Such a study is now in progress, and will be useful for the investigation of the realistic materials with strong correlation when it is combined with the band calculations.


\begin{thebibliography}{99}
\bibitem{Anderson61} P. W. Anderson: Phys. Rev. {\bf 124} (1961) 41.
\bibitem{Newns69} D. M. Newns: Phys. Rev. {\bf 178} (1969) 1123.
\bibitem{Goldhaber-Gordon98} See, for example, D. Golhaber-Gordon, H. Shtrikman, D. Mahalu, D. Abusch-Magder, U. Meirav and M. A. Kastner: Nature {\bf 391} (1998) 156.
\bibitem{Georges96} A. Georges, G. Kotliar, W. Krauth and M. J. Rozenberg: Rev. Mod. Phys. {\bf 68} (1996) 13.
\bibitem{Yoshida90} M. Yoshida, M. A. Whitaker and L. N. Oliveira: Phys. Rev. B {\bf 41} (1990) 9403.
\bibitem{Kuramoto83} Y. Kuramoto: Z. Phys. B {\bf 53} (1983) 37.
\bibitem{Grewe83} N. Grewe: Z. Phys. B {\bf 53} (1983) 271.
\bibitem{Coleman84} P. Coleman: Phys. Rev. B {\bf 29} (1984) 3035.
\bibitem{Read83} N. Read and D. M. Newns: J. Phys. C: Solid State Phys. {\bf 16} (1983) L1055.
\bibitem{Yamada75} K. Yamada: Prog. Theor. Phys. {\bf 53} (1975) 970.
\bibitem{Martin-Rodero82} A. Martin-Rodero, F. Flores, M. Baldo and R. Pucci: Solid State Commun. {\bf 44} (1982) 911.
\bibitem{Vilk94} Y. M. Vilk, L. Chen and A.-M. S. Tremblay: Phys. Rev. B {\bf 49} (1994) 13267.
\bibitem{White92} J. A. White: Phys. Rev. B {\bf 45} (1992) 1100.
\bibitem{Horvatic87} B. Horvati\'{c}, D. \v{S}ok\v{c}evi\'{c} and V. Zlati\'{c}: Phys. Rev. B {\bf 36} (1987) 675.
\bibitem{Bickers91} N. E. Bickers, D. J. Scalapino and S. R. White: Phys. Rev. Lett. {\bf 62} (1991) 961; N. E. Bickers and S. R. White: Phys. Rev. B {\bf 43} (1991) 8044.
\bibitem{Moriya85} T. Moriya: ''Spin fluctuations in Itinerant Eelectron Magnetism`` (Springer, 1985)
\bibitem{Moriya95} T. Moriya and T. Takimoto: J. Phys. Soc. Jpn. {\bf 64} (1995) 960.
\bibitem{Saso99} T. Saso: J. Phys. Soc. Jpn. {\bf 68} (1999) 3941.
\bibitem{Levy-Yeyati93} A. Levy Yeyati, A. Martin-Rodero and F. Flores: Phys. Rev. Lett. {\bf 71} (1993) 2991.
\bibitem{Takagi99} O. Takagi and T. Saso: J. Phys. Soc. Jpn. {\bf 68} (1999) 1997, 2894.
\bibitem{Kuramoto90} Y. Kuramoto and K. Miyake: J. Phys. Soc. Jpn. {\bf 59} (1990) 2831.
\bibitem{Okiji83} A. Okiji and N. Kawakami: Phys. Rev. Lett. {\bf 15} (1983) 1157.
\bibitem{Yamada79} K. Yamada: Prog. Theor. Phys. {\bf 62} (1979) 901.
\bibitem{Lange98} E. Lange: Mod. Phys. Lett. B {\bf 12} (1998) 915.
\end{thebibliography}
\end{document}